# gSmart: An Efficient SPARQL Query Engine Using Sparse Matrix Algebra – Full Version

Yuedan Chen
Hunan University
chenyuedan@hnu.edu.cn

M. Tamer Özsu
University of Waterloo
tamer.ozsu@uwaterloo.ca

Guoqing Xiao
Hunan University
gqxiao@hnu.edu.cn

Zhuo Tang
Hunan University
ztang@hnu.edu.cn

Kenli Li
Hunan University
lkl@hnu.edu.cn

## ABSTRACT

Efficient execution of SPARQL queries over large RDF datasets is a topic of considerable interest due to increased use of RDF to encode data. Most of this work has followed either relational or graph-based approaches. In this paper, we propose an alternative query engine, called gSmart, based on matrix algebra. This approach can potentially better exploit the computing power of high-performance heterogeneous architectures that we target. gSmart incorporates: (1) grouped incident edge-based SPARQL query evaluation, in which all unevaluated edges of a vertex are evaluated together using a series of matrix operations to fully utilize query constraints and narrow down the solution space; (2) a graph query planner that determines the order in which vertices in query graphs should be evaluated; (3) memory- and computation-efficient data structures including the light-weight sparse matrix (LSpM) storage for RDF data and the tree-based representation for evaluation results; (4) a multi-stage data partitioner to map the incident edge-based query evaluation into heterogeneous HPC architectures and develop multi-level parallelism; and (5) a parallel executor that uses the fine-grained processing scheme, pre-pruning technique, and tree-pruning technique to lower inter-node communication and enable high throughput. Evaluations of gSmart on a CPU+GPU HPC architecture show execution time speedups of up to 46920.00× compared to the existing SPARQL query engines on a single node machine. Additionally, gSmart on the Tianhe-1A supercomputer achieves a maximum speedup of 6.90× scaling from 2 to 16 CPU+GPU nodes.

## 1 INTRODUCTION

RDF is the standard and popular data model recommended by W3C that uses a set of triples ⟨subject, predicate, object⟩ to describe large-scale information in a wide range of domains, such as the semantic web (e.g., Linked Data [23]), knowledge graphs (e.g., Amazon and Google KGs), bioinformatics (e.g., Bio2RDF – bio2rdf.org), knowledge bases (e.g., Yago [39], DBpedia [7]). Each RDF triple describes the relationship (predicate) between subject and object (Figure 1a). RDF data can be modeled as an RDF graph, where the vertices represent subjects and objects, and the label of each edge from a subject vertex to an object vertex represents the corresponding predicate (Figure 1b).

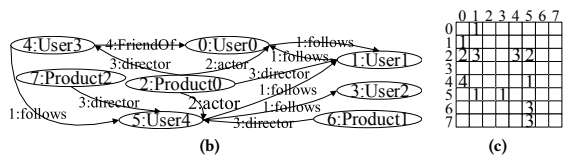

Figure 1: (a) RDF Triples; (b) RDF graph; (c) Corresponding sparse matrix.

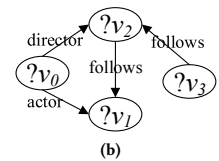

Figure 2: (a) SPARQL queries; (b) Corresponding query graph.

Efficiently executing RDF query language SPARQL has generated significant interest. Existing approaches to SPARQL query evaluation fall into two categories: relational or graph. The relational approaches are based on the join of the matches of each query triple to the RDF dataset and this matching is facilitated by elaborate index structures (e.g., [17, 30, 43]). The graph approaches convert SPARQL queries to a query graph, as shown in Figure 2, and evaluate queries by subgraph matching over the RDF graph (e.g., [2, 40, 47, 48]).

The increasing use of RDF in real applications has naturally been accompanied by a significant increase in the size of RDF datasets [17]. This has generated challenges in efficient SPARQL processing over large RDF datasets. Two main lines of research have focused on this problem. The first considers the use of hardware assists (GPUs, FPGAs) for query processing. The second focuses on scale-out architectures for parallel/distributed execution of SPARQL queries. While the scale-up versus scale-out processing is a topic of debate [29, 35], high performance computing (HPC) systems combine scale-up with scale-out through the use of many-/multi-core processors which have fine-grained parallel computing ability. HPC systems are more attuned for *matrix-based* computation (in contrast to relational and graph-based). The RDF graph can be represented as an $N \times N$ sparse matrix with $M$ nonzero entries corresponding to $M$ edges connecting the $N$ vertices in the RDF graph (Figure 1c). Each nonzero element $\{i, j\}$ in the sparse matrix with entry $l$ represents the predicate whose subject is $i$ and object is $j$ in RDF data with the $\overrightarrow{i, j}$ edge labelled $l$ representing the predicate.

This paper proposes gSmart, a matrix-based SPARQL query evaluation technique that is suitable for both heterogeneous architectures (consisting of CPU+GPU) and HPC architectures. The array-based and coarse-grained access patterns in the matrix-based paradigm can reduce some of the overhead of other approaches due to their unpredictable reference and element-wise access patterns that dominate query latency. Matrix-based processing may also be easier to parallellize and better match the memory hierarchy of HPC platforms [26]. We make two major contributions in this paper. First, we propose techniques to utilize heterogeneous CPU+GPU single-node systems to execute SPARQL systems. Second, we extend our solution to many-node, heterogeneous HPC systems that combines CPUs and many-/multi-core accelerators. In our experiments, we utilize the Tianhe-1A supercomputer (more information in §9), but the fundamental techniques and data structures apply to any heterogeneous HPC platform with massive number of CPU+GPU compute nodes.

The only matrix-based query engine that we are aware of is MAGiQ [25]. It translates each query edge in the SPARQL query graph to the customized language of matrix algebra with logical AND and OR operators and uses readily available libraries, i.e., SuiteSparse:GraphBLAS [13], Matlab, and CombBLAS [10], to process matrix algebra operations on the RDF matrix on both CPUs and GPUs. MAGiQ evaluates each query edge $\overrightarrow{v, w}$ by calculating sparse matrix operations using libraries and produces a binding matrix $M_{vw}$ of the same size as the RDF matrix, where the value "1" of element $\{i, j\}$ in $M_{vw}$ indicates that $i$ is the binding of variable $v$ and $j$ is the binding of variable $w$. However, MAGiQ's performance is limited by: (a) very heavy communication to access the large volume of intermediate results, and (b) arduous update calculations on the intermediate results to eliminate invalid bindings.

Implementing efficient matrix-based query evaluation on heterogeneous HPC platforms requires the solution of three main issues:

**C1: The memory footprint of the RDF matrix and binding matrices impact efficiency, both in terms of memory usage and communication.** The accelerator within each heterogeneous compute node can achieve the optimal performance only when the device memory contains the computational data. Therefore, the RDF matrix storage directly affects the GPU memory usage and communication overhead to load the memory. Furthermore, the intermediate results are usually returned from GPU device memory to CPU host memory in each compute node and even swapped between nodes for remaining query evaluation. Hence the storage of binding matrices determines the communication overhead.

**C2: The iterative update operations cause additional calculation and communication tasks.** The binding matrix gets updated as computation progresses, and if implemented naively, these updates cause additional computation. An example from how MAGiQ performs computation is illustrative. As MAGiQ successively evaluates query edges $\overrightarrow{v_0, v_1}$ and $\overrightarrow{v_1, v_2}$, some bindings of $v_1$ in binding matrix $M_{v_0 v_1}$ resulting from evaluating $\overrightarrow{v_0, v_1}$ may be invalidated by binding matrix $M_{v_1 v_2}$, which requires updating $M_{v_0 v_1}$ so that only bindings of $v_1$ stored in $M_{v_1 v_2}$ are included in updated $M_{v_0 v_1}$. For complex queries, the number of update operations is large, which adds extra computation and communication.

**C3: How to utilize the multi-stage and hybrid parallel computing architecture and memory hierarchy of heterogeneous HPC architectures for accelerating large-scale matrix-based RDF engine.** Massive heterogeneous compute nodes of HPC platforms form the process-level parallelism and parallel computing cores in accelerators form the thread-level parallelism. In addition, the memory hierarchy corresponds to the combination of host memory and device memory. To adapt to and exploit underlying heterogeneous HPC architectures, a multi-stage data partitioning scheme could be customized for the matrix-based query engine.

gSmart addresses these challenges within the context of a typical CPU+GPU heterogeneous architecture. It builds on the following key principles:

**Grouped incident edge-based query evaluation** (§ 5): We devise a grouped incident edge-based evaluation technique using matrix algebra primitives (operations). The technique fully utilizes constraints of incident edges of each vertex in query graphs to prune intermediate results and reduce update operations.

**Graph-based query planner** (§ 6.1): Incident edge-based query evaluation technique is sensitive to the processing order of query edges. gSmart adopts two traversal approaches to pick the order of query edges for processing to reduce the computation and communication volumes and increase the communication efficiency.

**Light-weight data structure** (§ 6.2 and § 7.1): gSmart uses the LSpM structure to store the input RDF data as a light-weight sparse matrix, where only necessary nonzeros for query evaluation are stored in the row- or/and column-wise fashion based on the query plan. This mitigates the device memory size limitation and reduces the intra-node communication overhead within each compute node. In addition, gSmart uses the tree-based binding storage to reduce memory footprint of variable bindings and the intra-/inter-node communication costs.

**Multi-stage data partitioner** (§ 6.3): gSmart adopts the multi-stage data partitioner that partitions RDF data in a fine-grained manner to leverage the multi-layer parallel computing architecture and surmount the limited device memory size and bandwidth. The multi-stage partitioner assigns exhaustive data for each compute node (based on the query graph structure and query plan) to avoid extra inter-node traffic.

**Efficient executor** (§ 7.2 and § 8): gSmart utilizes the heterogeneous execution pattern to fully utilize computing resources. Specifically, it incorporates a fine-grained query processing scheme and pre-pruning technique that together reduce the additional update calculation and communication problem highlighted above. In addition, we devise the local tree-pruning and main-process tree-pruning techniques to generate the final results (if required).

Our evaluation of gSmart, using both synthetic and real datasets, show superior performance and good scalability compared to other state-of-the-art techniques on heterogeneous architectures.

## 2 BASIC CONCEPTS

In this section we discuss three topics: (1) the fundamentals of matrix algebra, focusing on the operations that are important for this paper, (2) how SPARQL queries can be executed using these operators, and (3) CPU+GPU heterogeneous architectures that are the target of this paper.

### 2.1 Matrix Algebra

The matrix algebra operations include three groups: general selection operations, selections with predicates, and other operators.

*2.1.1 Selection Operations.* Row and column selection operations are in this category.

**Row selection.** Selecting the $i^{th}$ row of an $N \times N$ matrix $A$, called *row selection*, corresponds to multiplying a diagonal matrix $S$ with $A$, where only entry $\{i, i\}$ of $S$ ($S(i, i)$) is 1 and other entries are 0, and $i \in \{0, 1, 2, \ldots, N-1\}$.

**Column selection.** While selecting the $j^{th}$ column of an $N \times N$ matrix $A$, called *column selection*, corresponds to multiplying $A$ with a diagonal matrix $S$, where only entry $S(j, j)$ is 1 and other entries are 0, and $j \in \{0, 1, 2, \ldots, N-1\}$.

*Example 2.1.* A $3 \times 3$ matrix $A$ is presented as
$$A = \begin{bmatrix} a & b & c \\ d & e & f \\ g & h & i \end{bmatrix}. \quad (1)$$

Selecting the $0^{th}$ and $2^{nd}$ rows of $A$ can be calculated by
$$\begin{bmatrix} 1 & 0 & 0 \\ 0 & 0 & 0 \\ 0 & 0 & 1 \end{bmatrix} \times \begin{bmatrix} a & b & c \\ d & e & f \\ g & h & i \end{bmatrix} = \begin{bmatrix} a & b & c \\ 0 & 0 & 0 \\ g & h & i \end{bmatrix}. \quad (2)$$

Selecting the $2^{nd}$ column of $A$ can be calculated by
$$\begin{bmatrix} a & b & c \\ d & e & f \\ g & h & i \end{bmatrix} \times \begin{bmatrix} 0 & 0 & 0 \\ 0 & 0 & 0 \\ 0 & 0 & 1 \end{bmatrix} = \begin{bmatrix} 0 & 0 & c \\ 0 & 0 & f \\ 0 & 0 & i \end{bmatrix}. \quad (3)$$

*2.1.2 Predicate-Based Selection Operations.* We customize $\otimes$ as the operator for matrix multiplication, where the arithmetic multiplication operation "$\times$" and addition operation "$+$" are respectively replaced with logical AND "$\wedge$" and logical OR "$\vee$". The predicate-based selection operations are produced by matrix multiplications under the operator $\otimes$.

**Matrix-vector multiplication.** Finding row indices of predicate $p$ in matrix $A$ corresponds to multiplying $A$ with a vector $u_p$ under operator $\otimes$, denoted as $y = A \otimes u_p$, where all the entries of $u_p$ are $p$. This operation is expressed as
$$y(i) = \bigvee_{j} A(i, j) \wedge p, \quad (4)$$

where $y(i) = 1$ when $p$ exists in the $i^{th}$ row of $A$.

Finding column indices of predicate $p$ in $A$ corresponds to multiplying the transpose of $A$, denoted as $A^\top$, with vector $u_p$ under operator $\otimes$. This matrix-vector multiplication $y = A^\top \otimes u_p$ is expressed as
$$y(j) = \bigvee_{i} A(j, i) \wedge p, \quad (5)$$

where $y(j) = 1$ when $p$ exists in the $j^{th}$ column.

*Example 2.2.* Finding the rows that contain predicate $b$ in matrix $A$ shown in Eq. (1) can be expressed as
$$\begin{bmatrix} a & b & c \\ d & e & f \\ g & h & i \end{bmatrix} \otimes \begin{bmatrix} b \\ b \\ b \end{bmatrix} = \begin{bmatrix} 1 \\ 0 \\ 0 \end{bmatrix}. \quad (6)$$

Therefore, only the $0^{th}$ row of $A$ contains $b$.

Finding the columns that contain predicate $b$ in $A$ can be expressed as
$$\begin{bmatrix} a & b & c \\ d & e & f \\ g & h & i \end{bmatrix}^\top \otimes \begin{bmatrix} b \\ b \\ b \end{bmatrix} = \begin{bmatrix} a & d & g \\ b & e & h \\ c & f & i \end{bmatrix} \otimes \begin{bmatrix} b \\ b \\ b \end{bmatrix} = \begin{bmatrix} 0 \\ 1 \\ 0 \end{bmatrix}. \quad (7)$$

Therefore, only the $1^{st}$ column of $A$ contains $b$.

**Matrix-matrix multiplication.** Finding the positions of predicate $p$ in matrix $A$, including row indices and column indices, corresponds to multiplying a diagonal matrix $S_p$ with $A$ under operator $\otimes$, denoted as $M = S_p \otimes A$, where $S_p$ is a diagonal matrix that all entries on the main diagonal are $p$ and other entries are 0. This matrix-matrix multiplication is expressed as
$$M(i, j) = \bigvee_{k} S_p(i, k) \wedge A(k, j), \quad (8)$$

where $M(i, j) = 1$ means that $A(i, j) = p$.

*Example 2.3.* Selecting predicate $c$ in $A$ shown in Eq. (1) can be expressed as
$$\begin{bmatrix} c & 0 & 0 \\ 0 & c & 0 \\ 0 & 0 & c \end{bmatrix} \otimes \begin{bmatrix} a & b & c \\ d & e & f \\ g & h & i \end{bmatrix} = \begin{bmatrix} 0 & 0 & 1 \\ 0 & 0 & 0 \\ 0 & 0 & 0 \end{bmatrix}, \quad (9)$$

where $M(0, 2) = 1$ means that $A(0, 2) = c$.

*2.1.3 Other Operations.* We introduce vector AND and OR as part of our algebra.

**Vector AND.** This operation ($\odot$) is applied to two binary vectors of equal length, by taking the bitwise AND of each pair of elements at corresponding positions.

*Example 2.4.* Performing vector AND operation on a vector $x = \{1, 0, 1\}$ and another vector $y = \{0, 0, 1\}$ is presented as
$$\begin{bmatrix} 1 \\ 0 \\ 1 \end{bmatrix} \odot \begin{bmatrix} 0 \\ 0 \\ 1 \end{bmatrix} = \begin{bmatrix} 0 \\ 0 \\ 1 \end{bmatrix}. \quad (10)$$

**Vector OR.** This operation ($\oplus$) is applied to two binary vectors of equal length, by taking the bitwise OR of each pair of elements at corresponding positions.

*Example 2.5.* Performing vector OR operation on a vector $x = \{1, 0, 1\}$ and another vector $y = \{0, 0, 1\}$ is presented as

$$\begin{bmatrix} 1 \\ 0 \\ 1 \end{bmatrix} \oplus \begin{bmatrix} 0 \\ 0 \\ 1 \end{bmatrix} = \begin{bmatrix} 1 \\ 0 \\ 1 \end{bmatrix}. \quad (11)$$

## 2.2 Matrix Algebra-based Query Processing

*2.2.1 RDF and SPARQL Basics.* As noted earlier, RDF models each fact as a triple ⟨**s**ubject, **p**redicate, **o**bject⟩, denoted as ⟨$s, p, o$⟩, where *subject* is an entity, class or blank node, a *predicate* denotes one attribute associated with one entity, and *object* is an entity, a class, a blank node, or a literal value. According to the RDF standard, an entity is denoted by a URI (Uniform Resource Identifier) that refers to a named *resource* in the environment that is being modeled. Blank nodes, by contrast, refer to anonymous resources that do not have a name. Thus, each triple represents a named relationship; those involving blank nodes simply indicate that "something with the given relationship exists, without naming it" [27].

Formally, if $\mathcal{U}, \mathcal{B}, \mathcal{L}$, and $\mathcal{V}$ denote the sets of all URIs, blank nodes, literals, and variables, respectively, a tuple $(s, p, o) \in (\mathcal{U} \cup \mathcal{B}) \times \mathcal{U} \times (\mathcal{U} \cup \mathcal{B} \cup \mathcal{L})$ is an *RDF triple*. A set of RDF triples form a *RDF data set*.

As noted earlier, RDF data can be modeled as an RDF graph where there is a vertex corresponding to each unique subject and object, and a directed edge $\overrightarrow{v_i, v_j}$ with label $l$ connecting vertices $v_i$ and $v_j$ exists if there is corresponding triple in the RDF data set.

A SPARQL query can be defined formally as follows. Let $\mathcal{U}, \mathcal{B}, \mathcal{L}$, and $\mathcal{V}$ denote the sets of all URIs, blank nodes, literals, and variables, respectively. A SPARQL expression is expressed recursively

- A *triple pattern* $(\mathcal{U} \cup \mathcal{B} \cup \mathcal{V}) \times (\mathcal{U} \cup \mathcal{V}) \times (\mathcal{U} \cup \mathcal{B} \cup \mathcal{L} \cup \mathcal{V})$ is a SPARQL expression,
- (optionally) If $P$ is a SPARQL expression, then $P$ FILTER $R$ is also a SPARQL expression where $R$ is a built-in SPARQL filter condition,
- (optionally) If $P_1$ & $P_2$ are SPARQL expressions, then $P_1$ AND| OPT|OR $P_2$ are also SPARQL expressions.

A set of triple patterns is called *basic graph pattern* (BGP) and SPARQL expressions that only contain these are called *BGP queries*. These are the subjects of most of the research in SPARQL query evaluation.

A SPARQL query can also be represented as a *query graph* similar to the RDF mapping: each vertex corresponds to a subject, object or variable and each edge corresponds to a predicate. An edge with the two end points is a triple pattern.

*2.2.2 Evaluation on a Single Query Edge.* Evaluation of a single query edge corresponds to evaluating a triple pattern. Consider the following query (assume the degree of vertices $x$ and $y$ are 1) that asks for bindings of variables $x$ and $y$ that satisfy the triple pattern $?x \, p_{xy} \, ?y$ – i.e., it is looking for triples involving variables $x$ and $y$ that with predicate $p_{xy}$:
SELECT ?x ?y WHERE {?x $p_{xy}$ ?y .}

The query evaluation can be translated to a predicate-based selection operation that finds the positions of predicate $p_{xy}$ in RDF matrix $A$, i.e., multiplying diagonal matrix $S_{p_{xy}}$ with $A$ under the operator $\otimes$, where all entries on the main diagonal are $p_{xy}$:

$$M_{xy} = S_{p_{xy}} \otimes A = p_{xy} \times I \otimes A, \quad (12)$$

where $I$ is an identity matrix. Result matrix $M_{xy}$ stores bindings of variables $x$ and $y$ (called the *binding matrix*), where $M_{xy}(i, j) = 1$ means that $i$ and $j$ are bindings of subject $x$ and object $y$, respectively. Elementwise, $M_{xy}(i, j)$ is calculated by

$$\begin{aligned} M_{xy}(i, j) &= \bigvee_k S_{p_{xy}}(i, k) \wedge A(k, j) \\ &= S_{p_{xy}}(i, i) \wedge A(i, j) = p_{xy} \wedge A(i, j). \end{aligned} \quad (13)$$

Therefore, the result matrix $M_{xy}$ is calculated by executing logical OR operation on predicate $p_{xy}$ and each nonzero of $A$.

*2.2.3 Evaluation on Conjunctive Query Graphs.* When a SPARQL query involves multiple triple patterns, its semantics is the conjunction of these triple patterns. Such a query can be evaluated using matrix algebra operators as follows. Given a binding matrix $M_{xy}$ extracted from a RDF matrix $A$, assume there is a query edge $\overrightarrow{y, z}$ with predicate $p_{yz}$, and there are no other edges of vertex $y$ to be evaluated. There are two steps of the query evaluation:

**Step 1**: Compute bindings of variable $y$.

The bindings of $y$ can be transformed into a binding vector $v_y$:

$$v_y = \bigoplus_i M_{xy}^\top(:, i), \quad (14)$$

where $M_{xy}^\top(:, i)$ is the vector of $i^{th}$ column of $M_{xy}^\top$, i.e., the $i^{th}$ row vector of $M_{xy}$.

**Step 2**: Compute the binding matrix $M_{yz}$.

According to Eq. (12) and Eq. (14), the binding matrix $M_{yz}$ can be computed by:

$$M_{yz} = p_{yz} \times I \otimes (diag(v_y) \times A) \quad (15)$$

where $diag(\cdot)$ is a function that outputs a diagonal matrix $S$ where each entry $S(i, i)$ on the main diagonal is corresponding entry $v_i$ of input vector $v$. Elementwise, $M_{yz}$ is calculated by

$$M_{yz}(i, j) = p_{yz} \wedge A(i, j), \left\{ i \, \middle| \, \bigvee_j M_{xy}(j, i) \neq 0 \right\}. \quad (16)$$

## 2.3 CPU+GPU Heterogeneous Architecture

GPUs have been used as coprocessors with CPUs to provide high throughput for data-intensive applications. The Computing Unified Device Architecture (CUDA) is created by NVIDIA to enable general purpose processing on GPUs, termed general-purpose GPUs (GPGPUs), and improve the efficiency of parallel program development. The basic parallel computing unit of a GPU is a Stream Multiprocessor (SM). Each SM consists of a set of processors, named as SPs (Stream Processors), where the parallel computing cores in each SP are arranged as an array, thus supporting thousands of concurrent threads. Each SM concurrently executes the same instruction on a group of threads, referred to as *wrap*, within a clock cycle. In addition, CUDA threads are extremely lightweight, wherein the GPU manages and controls the threads with very little overhead.

Figure 3 presents the heterogeneous parallel computing system of a CPU+GPU machine. There are two portions of tasks on the heterogeneous architecture. One portion is executed on the GPU (the *device*), utilizing powerful parallel computing resources with a number of concurrent CUDA threads. Another portion is executed on the CPU (the *host*), generally including management of GPU tasks, data swapping, and performing computations that are not suitable for executing on the GPU. The heterogeneous parallel computing pattern mainly consists of three parts. First, the host allocates spaces in device memory and transfers computational data from host memory to device memory via the PCIe (Peripheral Component Interconnect Express) bus. Second, the device executes kernels in parallel. Third, results are returned from device memory to host memory via PCIe.

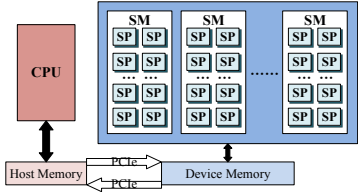

Figure 3: A CPU+GPU architecture.

A typical CPU+GPU heterogeneous HPC platform is equipped with a large number CPU+GPU compute nodes, which provide the process-stage parallelism for the platform. Furthermore, parallel computing cores of the GPU within each compute node provide the finer-grained thread-stage parallelism. Accelerating applications on heterogeneous HPC platforms mainly face three challenges:

(1) The elaborate development of multi-stage and heterogeneous parallel computing architecture;
(2) The full utilization of relatively small capacity of device memory and limited PCIe bandwidth of intra-node communication in each compute node;
(3) The alleviation of costly inter-node communication between hosts of compute nodes.

In this paper we discuss how gSmart addresses these challenges.

## 3 RELATED WORK

There has been significant research on optimizing RDF query engines that run on a single machine [8, 37, 42, 46, 48]. OBDA [37] handles RDF data directly using a relational database technology. IBM's DB2RDF [8] denormalizes the triples table into clustered properties. gStore [48] and GQA$_{RDF}$ [42] execute subgraph matching on the RDF graph to evaluate queries.

As RDF data sizes have increased, research has focused on designing distributed query engines. Some of the distributed RDF query engines utilize computing frameworks such as Spark [6, 12, 36], MapReduce [15, 31, 34], and Pregel [41, 44, 45]. Others include the engines utilizing specific physical structures, inherent RDF indexing, high performance communication mechanisms, query optimizations [3, 11, 17, 18, 40, 47], etc.

There has been some work in improving the performance of database operations using heterogeneous architectures such as CPU+GPU [19–21, 32]. HyPE [9] decides the optimal execution based on the estimated runtimes for database management systems on GPU. He *et al.* [22] develop a cost model-guided adaptation mechanism for distributing workloads between CPU and GPU, and use device fission to divide the CPU or the GPU into fine-grained units for better resource utilization. SABER [28] executes window-based relational stream processing and increases the share of queries executing on CPU+GPU architectures to yield the highest performance based on the past behaviour. GPL [33] uses an analytical model to generate the optimal pipelined query plan, so that the tile size of the pipelined query execution can be adapted in a cost-based manner. Based on the pipelined query plan generated by GPL, Pyper [32] provides the *Shuffle* operator for a pipeline to reduce divergent execution and the *Segment* operator that splits a workload-overwhelmed pipeline for higher thread parallelism, for a just-in-time compile-based query engine on GPU. Doraiswamy and Freire [14] present a geometric data model that provides a uniform geometric representation for different spatial data objects, and design an algebra consisting of GPU-friendly composable operators that can handle spatial queries. To the best of our knowledge, none of the above systems was developed for exploiting a distributed heterogeneous CPU+GPU environment.

Wukong+G [40] exploits a heterogeneous CPU+GPU cluster for graph exploration-based SPARQL query processing by designing the GPU-based query execution to surmount the limitations of GPU memory size and PCIe (Peripheral Component Interconnect Express) bandwidth, a GPU-friendly RDF store to aggregate keys and values with the same predicate individually and cache them in GPU memory, and a heterogeneous communication framework to transfer metadata among CPUs via native RDMA and intermediate results among GPUs via GPUDirect RDMA. However, fine-grained irregular data accessing patterns in graph-based engines limit the utilization of memory bandwidth and parallelism, which can be mitigated by matrix algebra-based query evaluation.

MAGiQ [25] is closest to our work in that it uses a matrix-based approach for processing SPARQL queries over a heterogeneous architecture. It uses existing libraries to evaluate matrix-based SPARQL queries. However, MAGiQ shows no significant performance advantage because optimizations, including data structure, query planner, data partitioning, efficient parallel execution, etc., are not studied in detail to exploit the computing power of underlying platforms. In contrast, gSmart provides an end-to-end solution and exploits heterogeneous HPC platforms to further accelerate matrix algebra-based RDF engine by devising the grouped incident edge-based query evaluation, graph-based query planer, computing- and memory-efficient data storage, multi-stage data partitioner, and efficient parallel executor.

## 4 GSMART – OVERVIEW

gSmart involves three computation phases: pre-processing, main computation, and post-processing. Figure 4 shows the overview of gSmart execution on a CPU+GPU heterogeneous architecture.

*Pre-processing phase.* The pre-processing phase is executed on the CPU of each compute node.

A *graph-based query planner* is used to parse the SPARQL query to generate a query graph and pick the processing order of query edges using graph traversal methods. The direction of each query edge determines whether to select rows or columns in the RDF matrix or binding matrix for evaluation. The degree of each query vertex affects the utilization of constraints in query evaluation. Therefore, we propose two approaches to pick the order of query

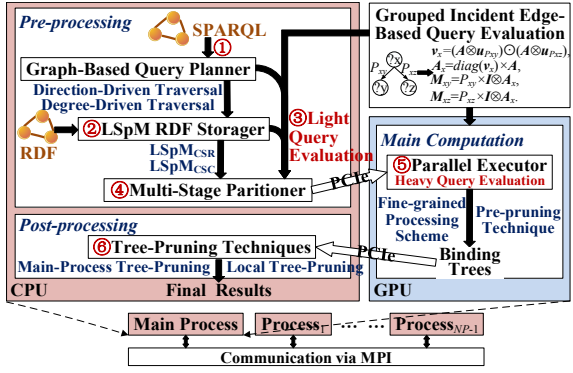

Figure 4: The overview of gSmart on a CPU+GPU heterogeneous HPC architecture.

edges for evaluation: direction-driven traversal and degree-driven traversal (§ 6.1).

gSmart uses a *LSpM RDF storage system* to reduce the volume of data swapping between the host memory and device memory at each compute node. LSpM only stores the necessary RDF triples containing the predicates that occur in the query graph in a compressed sparse matrix, based on the processing order and directions of query edges (§ 6.2).

gSmart uses a *multi-stage partitioner* that partitions the LSpM-stored RDF matrix to adapt to the multi-stage parallel computing architecture and memory structure of the platform. Additionally, the multi-stage data partitioner assigns required rows or/and columns in the LSpM-stored matrix to compute nodes, according to the evaluation direction and the predicate of each query edge, to reduce data exchange between them (§ 6.3).

During pre-processing, we also consider the query structure with cycles and constants in making decisions. Specifically, gSmart processes query edges involving constant vertices, which we refer as *light queries*, on CPUs before multi-stage partitioning.

*Main computation phase.* The main computation phase corresponds to evaluating heavy queries on GPU threads. *Heavy queries* refer to query edges that only contain variable vertices.

gSmart utilizes a *tree-based binding structure* to compress the storage of variable bindings. The tree-based binding storage allows the flexibility of indexing arbitrary query results, which is memory- and computing-friendly (§ 7.1).

In addition, gSmart follows a *fine-grained processing scheme* where pruning takes place on each compute node to load LSpM-stored RDF data from host memory to device memory, execute the grouped incident edge-based query evaluation for heavy queries on the GPU, and return the binding trees from the device memory to host memory (§ 7.2).

*Post-processing phase.* The post-processing phase updates the binding trees using the *local tree-pruning technique* or/and *main-process tree-pruning technique* on CPUs to obtain the final query results according to the query graph structure. The local tree-pruning is performed on the local CPU of each compute node, while the main-process tree-pruning is executed on the CPU of the master compute node (§ 8).

*Example 4.1.* Taking the SPARQL queries shown in Figure 2 as example, as shown in Figure 4, each CPU first finds the processing order of edges in the query graph (①) and stores the input RDF data in LSpM format (②). In this example, gSmart skips the light query evaluation (③), since there are no constant vertices in the given query graph and all the edges are heavy queries instead of light queries. Next, the CPU partitions the LSpM-stored RDF matrix (④), and evaluates all the edges in the query graph (heavy queries) by leveraging all the GPU threads (⑤). GPU returns the evaluation results stored in binding trees. According to the processing order of edges and the query graph structure, the binding trees are pruned to compute the final result (⑥).

## 5 GROUPED INCIDENT EDGE-BASED QUERY EVALUATION

An important aspect of the gSmart solution is that all the unevaluated query edges incident to a vertex are evaluated together to restrict bindings of the vertex through the use of the edge predicates. We call this *grouped incident edge-based evaluation* and it helps prune invalid results using sparse matrix operations. This computation is performed using the matrix algebra operators introduced earlier. We introduce the technique in this section as it is used in the subsequent discussions.

Consider the following SPARQL query (assume the degree of vertex $x$ is 2, where the indegree is 0 and the outdegree is 2) that looks for triples involving variables $x$ and $y$ with predicate $p_{xy}$, and triples involving variables $x$ and $z$ with predicate $p_{xz}$:

SELECT ?x ?y ?z WHERE {?x $p_{xy}$ ?y. ?x $p_{xz}$ ?z.}

We utilize the two constraints of the outgoing query edges $\overrightarrow{x,y}$ and $\overrightarrow{x,z}$ to narrow down the solution space of variable $x$. As shown in Figure 5, there are three steps of the query evaluation:

**Step 1**: *Compute bindings of variable $x$ that satisfy constraints of the two query edges.*

This step corresponds to finding rows in the RDF matrix $A$ that contain both predicates $p_{xy}$ and $p_{xz}$. This is done as follow:

$$v_x = (A \otimes u_{p_{xy}}) \odot (A \otimes u_{p_{xz}}) \quad (17)$$

where $u_{p_{xy}}$ and $u_{p_{xz}}$ are vectors where all the elements are $p_{xy}$ and $p_{xz}$, respectively. $v_x$ is the binding vector of $x$ where $v_x(i) = 1$ means that $i$ is the binding of $x$.

**Step 2**: *Select rows of $A$, where the row indices correspond to bindings of $x$, based on $v_x$.*

This is done as follow:

$$A_x = S_x \times A = diag(v_x) \times A. \quad (18)$$

**Step 3**: *Compute the binding matrices $M_{xy}$ and $M_{xz}$ based on $A_x$.*

This step corresponds to finding the positions of $p_{xy}$ and $p_{xz}$ in matrix $A_x$. Hence, we can obtain $M_{xy}$ and $M_{xz}$ by

$$M_{xy} = p_{xy} \times I \otimes A_x, M_{xz} = p_{xz} \times I \otimes A_x. \quad (19)$$

Elementwise, $M_{xy}$ and $M_{xz}$ are calculated by

$$M_{xy}(i,j) = p_{xy} \wedge A(i,j), \ M_{xz}(i,k) = p_{xz} \wedge A(i,k),$$
$$\left\{ i \middle| \left(\bigvee_j p_{xy} \wedge A(i,j)\right) \wedge \left(\bigvee_k p_{xz} \wedge A(i,k)\right) \neq 0 \right\}. \quad (20)$$

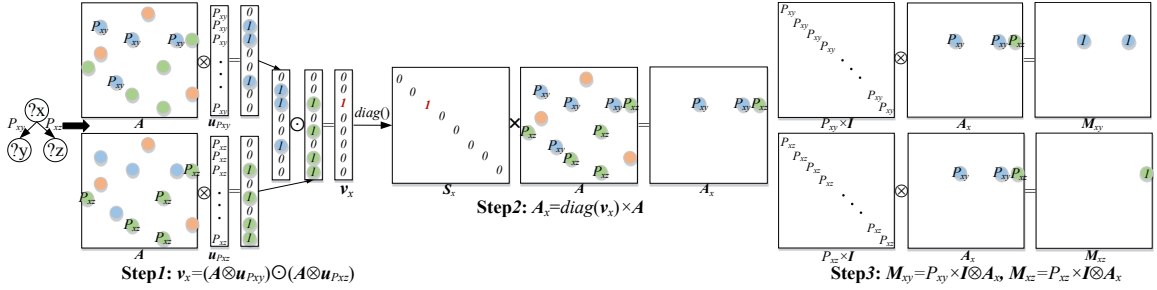

Figure 5: An example of grouped incident edge-based query evaluation.

Consider the following SPARQL query (assume the indegree of vertex x is 1 and the outdegree is 1) that looks for triples involving variables $x$ and $y$ with predicate $p_{yx}$, and triples involving variables $x$ and $z$ with predicate $p_{xz}$:

SELECT ?x ?y ?z WHERE {?y $p_{yx}$ ?x. ?x $p_{xz}$ ?z.}

The evaluation of this query requires two changes to the evaluation on the incident outgoing query edges of $x$:

At **Step 1**, the binding vector of $x$ is computed as:
$$v_x = (A^\top \otimes u_{p_{yx}}) \odot (A \otimes u_{p_{xz}}). \quad (21)$$

At **Step 2**, based on $v_x$, rows and columns of $A$ are selected where the row and column ids correspond to bindings of $x$. The row selection and column selection are:
$$A_x^r = S_x \times A = diag(v_x) \times A, \; A_x^c = A \times S_x = A \times diag(v_x). \quad (22)$$

At **Step 3**, the binding matrices $M_{xz}$ and $M_{yx}$ are respectively calculated as
$$M_{xz} = p_{xz} \times I \otimes A_x^r, M_{yx} = p_{yx} \times I \otimes A_x^c. \quad (23)$$

Elementwise, $M_{xz}$ and $M_{yx}$ are calculated by
$$M_{xz}(i,j) = p_{xz} \wedge A(i,j), M_{yx}(k,i) = p_{yx} \wedge A(k,i),$$
$$\left\{ i \middle| \left(\bigvee_j p_{xz} \wedge A(i,j)\right) \wedge \left(\bigvee_k p_{yx} \wedge A(k,i)\right) \neq 0 \right\}. \quad (24)$$

## 6 PRE-PROCESSING PHASE

This section introduces three key techniques in the pre-processing phase of gSmart, including graph-based query planner, LSpM RDF storage, and multi-stage data partitioner.

### 6.1 Graph-Based Query Planner

gSmart first parses the SPARQL query into a query graph (Figure 2). The direction of each query edge determines whether to evaluate on selected rows or columns in the RDF matrix or binding matrices. The degree of each query vertex affects the utilization of constraints in incident edge-based query evaluation. To determine the order of evaluation of the edges in the query graph, we propose two traversal approaches: direction-driven traversal and degree-driven traversal.

*6.1.1 Direction-driven Traversal.* The direction-driven traversal finds the order of query edges using DFS, where each path is explored along edge directions as deeply as possible before backtracking. gSmart processes all the unevaluated outgoing edges of a vertex $v$ together once $v$ is visited (as discussed in § 5). Given a query graph $G_q = (E, V)$, the detailed workflow of direction-driven traversal is as follows:

(1) Initialize a stack $S = \varnothing$, sets $W = \varnothing$ and $F = \varnothing$, and $r = 0$. $S$ is used to remember to get the next vertex to start a search when a dead-end occurs in any iteration of the traversal, $W$ ($F$) holds the vertices (edges) that have so far been visited (evaluated), and $r$ counts the number of traversal roots.
(2) Find a vertex in $V - W$ that has no unevaluated incoming edges as the $r^{th}$ root ($Root_r$) to start. If more than one vertex satisfies this condition, select one that has the maximum number of unevaluated outgoing edges. Push $Root_r$ onto $S$.
(3) If $S \neq \varnothing$, pop a vertex $v$ from $S$; Otherwise, skip to step 5.
(4) Evaluate all the unevaluated outgoing edges of $v$ (i.e., $\{\overrightarrow{v, w} \in E - F\}$). Add the set of edges $\{\overrightarrow{v, w}\}$ to $F$, add set of vertices $\{w\}$ to $W$, and push $\{w\}$ onto $S$ in ascending order of the number of unevaluated outgoing edges. Set $i = i + 1$, and go back to step 3.
(5) If $E - F = \varnothing$, end. Otherwise, set $r = r + 1$ and go back to step 2 to find a new root.

Based on direction-driven traversal, the order of each edge for evaluation is consistent with the corresponding edge direction, so that the query evaluation only accesses required rows of $A$.

If $G_q$ is **cyclic**, then step 2 is changed to:

(2) Find a vertex in $V - W$ that has no unevaluated incoming edges as $Root_r$ to start. If more than one vertex satisfies this condition, select one that has the maximum number of unevaluated outgoing edges. If no vertex satisfies this condition, find one in $V - W$ that has the maximum number of unevaluated outgoing edges. Push $Root_r$ onto $S$.

If $G_q$ has **constant** vertices, the processing order of query edges is obtained by the degree-driven traversal that is discussed below.

*Example 6.1.* Consider the query graph in Figure 2b. The evaluation order of all edges using direction-driven traversal is: $\{\overrightarrow{v_0, v_1}, \overrightarrow{v_0, v_2}\}$, $\{\overrightarrow{v_2, v_1}\}$, $\{\overrightarrow{v_3, v_2}\}$, where $Root_0$ is $v_0$ and $Root_1$ is $v_3$.

*6.1.2 Degree-driven Traversal.* The degree-driven traversal picks the order of query edges using DFS, where each path is searched as deeply as possible before backtracking, regardless of edge direction. In particular, all the unprocessed incident edges of a vertex $v$ are evaluated after $v$ is visited. The detailed workflow is as follows:

(1) Initialize a stack $S = \varnothing$, sets $W = \varnothing$ and $F = \varnothing$, and $r = 0$. These have the same meanings as in direction-driven traversal.
(2) Find a vertex that has the maximum number of unevaluated edges (incoming and outgoing) in $V - W$ as $Root_r$ to start. If more than one vertex satisfies this condition, select one that has the maximum number of unevaluated outgoing edges. Push $Root_r$ onto $S$.
(3) If $S \neq \varnothing$, pop a vertex $v$ from $S$; Otherwise, skip to step 5.
(4) Evaluate all the unevaluated edges of $v$ (i.e., $\{\overrightarrow{v,w} \in E - F\}$ and $\{\overrightarrow{w,v} \in E - F\}$). Then, add $\{\overrightarrow{v,w}\}$ and $\{\overrightarrow{w,v}\}$ to $F$, add $w$ to $W$, and push $w$ onto $S$ in ascending order of the number of unevaluated edges (or in ascending order of the number of unevaluated outgoing edges if the number of unevaluated edges is the same). Set $i = i + 1$, and go back to step 3.
(5) If $E - F = \varnothing$, end. Otherwise, set $r = r + 1$ and go back to step 2 to find a new root.

Based on degree-driven traversal, the evaluation order of partial vertices is inconsistent with the directions of corresponding edges, so that the query evaluation may access required rows and columns of $A$.

If $G_q$ is **cyclic**, the traversal method remains unchanged. If $G_q$ has **constants**, steps 1 and 2 of the degree-driven traversal method are changed to:

(1) Initialize a stack $S = \varnothing$, a set $F = \varnothing$, and $r = 0$; Add all the constant vertices into set $W$, evaluate all the incident edges of constant vertices, and add these edges to $F$.
(2) Find a vertex that has the maximum number of unevaluated edges in the adjacent vertices of the constants as $Root_r$ to start. If more than one vertex satisfies this condition, select one that has the maximum number of unevaluated outgoing edges. Push $Root_r$ onto $S$.

*Example 6.2.* Considering the query graph in Figure 2b, the processing order of edges using degree-driven traversal is: $\{\overrightarrow{v_0,v_2}, \overrightarrow{v_2,v_1}, \overrightarrow{v_3,v_2}\}$, $\{\overrightarrow{v_0,v_1}\}$, where $Root_0$ is $v_2$.

## 6.2 Light-weight Sparse Matrix (LSpM) RDF Storage

To reduce the memory footprint and communication, we propose the LSpM RDF storage system to store RDF data as a sparse matrix based on the evaluation order of query edges as determined in § 6.1.

### 6.2.1 Direction-driven LSpM.
The LSpM RDF structure based on the direction-driven query plan stores the RDF matrix by rows (row-wise LSpM$_{CSR}$ format). There are four steps for storing RDF data with $M$ triples:

(1) Read necessary RDF triples where predicates appear in the queries.
(2) Encode RDF strings into numeric ids following the common practice [4], where the index of subject and object is 0-based, the index of predicate is 1-based. Present the corresponding RDF matrix $A$.
(3) Eliminate empty rows of $A$.
(4) Store the row-wise reduced $A$ in CSR format, named the LSpM$_{CSR}$ format, where the nonzeros are stored by rows.

*Example 6.3.* Consider the RDF data in Figure 1b and the SPARQL queries in Figure 2. Step 1 deletes RDF triples with predicate FriendOf (which does not appear in the queries). Step 2 establishes the corresponding RDF matrix $A$ having 8 rows, 8 columns, and 11 nonzeros. Step 3 computes the reduced $A$ with 7 rows, and an array $Mr[9] = \{0, 1, 2, 3, 3, 4, 5, 6, 7\}$ is built to mark the elimination of each row, where $Mr[i + 1] - Mr[i] = 1$ indicating that the $i^{th}$ row of $A$ is non-empty and this row is the $Mr[i]^{th}$ row in the reduced $A$. Step 4 stores the reduced $A$ using three LSpM$_{CSR}$ arrays: $Pr[8] = \{0, 1, 2, 6, 7, 9, 10, 11\}$ stores the pointers to the start and end positions of each non-empty row, $Val[11] = \{1, 1, 2, 3, 3, 2, 1, 1, 1, 3, 3\}$ stores the numerical value of each nonzero, and $Col[11] = \{1, 0, 0, 1, 4, 5, 0, 5, 1, 5, 5\}$ stores the column index of each nonzero.

### 6.2.2 Degree-driven LSpM.
The LSpM RDF structure based on the degree-driven query plan stores $A$ by rows or/and columns. For the edges whose directions are consistent with the evaluation order, the LSpM structure stores the corresponding data of $A$ by rows (row-wise LSpM$_{CSR}$ format). For edges whose directions are opposite to the evaluation order, the LSpM structure stores the corresponding data of $A$ by columns (column-wise LSpM$_{CSC}$ format).

There are five steps for storing the RDF data with $M$ triples in LSpM (row-wise or/and column-wise):

(1) Read necessary RDF triples where predicates appear in the query.
(2) Encode RDF strings into numeric ids and present the corresponding RDF matrix $A$.

*Row-wise LSpM$_{CSR}$ format*:

(3) Eliminate nonzeros in $A$ where corresponding RDF predicates do not appear in direction-consistent query edges.
(4) Eliminate empty rows of $A$.
(5) Store the row-wise reduced $A$ in CSR format.

*Column-wise LSpM$_{CSC}$ format*:

(3) Eliminate nonzeros in $A$ where corresponding RDF predicates do not appear in direction-opposite query edges.
(4) Eliminate empty columns of $A$.
(5) Store the column-wise reduced $A$ in CSC format, named the LSpM$_{CSC}$ format, where the nonzeros are stored by columns.

If the query contains **constants**, the outgoing edges of constants are included in direction-consistent queries, and the incoming edges of constants are included in direction-opposite queries.

*Example 6.4.* Considering the RDF data in Figure 1b and the SPARQL query in Figure 2b when the degree-driven traversal is used, the predicates that do not appear in the direction-consistent query edges ($\overrightarrow{v_2,v_1}$ and $\overrightarrow{v_0,v_1}$) are director and FriendOf, and the predicates that do not appear in direction-opposite edges ($\overrightarrow{v_0,v_2}$ and $\overrightarrow{v_3,v_2}$) are actor and FriendOf.

Steps 1 and 2 delete RDF triples with predicate FriendOf and generate the corresponding RDF matrix $A$.

As shown in Figure 6, to store $A$ in LSpM$_{CSR}$ format, step 3 further eliminates predicate director in $A$. Step 4 computes the reduced $A$ having 5 rows, 8 columns, and 7 noneros. The array $Mr[9]$ is built to mark the elimination of each row. Step 5 stores the reduced $A$ in LSpM$_{CSR}$ format using three arrays: $Pr[6]$, $Val[7]$, and $Col[7]$, to respectively store the pointers to non-empty rows, the numerical value and column index of each nonzero.

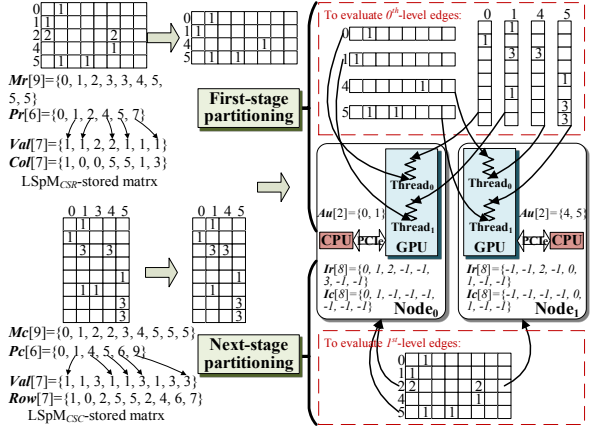

Figure 6: An example of the multi-stage partitioner of parallel degree-driven gSmart, where the example RDF data is shown in Figure 1b and queries are shown in Figure 2.

As shown in Figure 6, to store $A$ in $LSpM_{CSC}$ format, step 3 further eliminates predicate actor in $A$. Step 4 computes the reduced $A$ having 8 rows, 5 columns, and 9 nonzeros. The array $Mc[9]$ is built to mark the elimination of each column. Step 5 stores the reduced $A$ in $LSpM_{CSC}$ format using three arrays: $Pc[6]$, $Val[9]$, and $Row[9]$, to respectively store the pointers to non-empty columns, the numerical value and row index of each nonzero.

## 6.3 Multi-Stage Data Partitioner

In gSmart, each GPU thread evaluates the entire query graph on the partial RDF data the GPU receives. In preparation, the RDF matrix is partitioned using a technique that utilizes the number of edge levels from each traversal root – hence the term multi-stage data partitioning. The level of the edge $\overrightarrow{v,w}$ visited from the $Root_r$ is defined as the maximum distance from $Root_r$ to $v$ and $w$. Let $L_r$ be the number of levels of all the query edges visited from each $Root_r$, then the maximum number of edge levels of all the roots is $L = \max\{L_r\}$.

*6.3.1 Direction-driven Partitioner.* Based on the order in which query edges are picked, the multi-stage RDF partitioner divides the matrix $A$ stored in $LSpM_{CSR}$ format by rows.
**First-stage partitioning:** To evaluate the $0^{th}$-level query edges of all the roots, partition $A$ into $N_p \times N_t$ parts based on the number of rows for all the $N_p \times N_t$ GPU threads (where $N_p$ is the number of processes, corresponding to compute nodes, and $N_t$ is the number of GPU threads launched on each node).

Each compute node holds $N_t$ parts, and each GPU thread in the node evaluates the $0^{th}$-level query edges on one part, individually.
**Next-stage partitioning**: To evaluate the $l^{th}$-level query edges ($l \in \{1, 2, \ldots, L-1\}$), each node also holds the other rows in $A$ that could be used for the evaluation. The indices of these rows are the column indices of the nonzeros in the rows used for evaluating the $(l-1)^{th}$-level query edges.

Based on the direction-driven query plan, the multi-stage data partitioner for **cyclic** queries remains unchanged.

*6.3.2 Degree-driven Partitioner.* Based on the order in which query edges are picked, the multi-stage RDF partitioner partitions the $LSpM_{CSR}$-stored $A$ by rows and the $LSpM_{CSC}$-stored $A$ by columns.
**First-stage partitioning**: To evaluate the $0^{th}$-level query edges of all the roots, the first-stage partitioning varies according to the direction of the $0^{th}$-level query edges:

- If they are all direction-consistent, the first-stage partitioning is the same as the direction-driven partitioner.
- If they are all direction-opposite, the $LSpM_{CSC}$-stored $A$ is divided by columns into $N_p \times N_t$ parts based on the number of columns. Each compute node holds $N_t$ parts and evaluates the $0^{th}$-level query edges on them.
- If they are both direction-consistent and direction-opposite, delete the $Mr[k]^{th}$ rows in the $LSpM_{CSR}$-stored $A$ if the $k^{th}$ column of $A$ does not exist in the $LSpM_{CSC}$-stored $A$, and delete the $Mc[k]^{th}$ column in the $LSpM_{CSC}$-stored $A$ if the $k^{th}$ row of $A$ does not exist in $LSpM_{CSR}$-$A$; Thus the rows and columns that have the same indices in $A$ are retained. Then partition the $LSpM_{CSR}$-stored $A$ and the $LSpM_{CSC}$-stored $A$ into $N_p \times N_t$ parts, where each part has rows and columns that have the same indices in $A$. The direction-consistent edges are evaluated on the rows in the parts, and the direction-opposite edges are evaluated on the columns in the parts.

**Next-stage partitioning**: To evaluate the $l^{th}$-level query edges ($l \in \{1, 2, \ldots, L-1\}$), each node also holds the other rows in the $LSpM_{CSR}$-stored $A$ required for evaluating direction-consistent edges and other columns in the $LSpM_{CSC}$-stored $A$ required for evaluating direction-opposite edges.

The next-stage partitioning for evaluating the $l^{th}$-level edges varies according to the direction of the $(l-1)^{st}$-level query edges that have the common vertices with the $l^{th}$-level edges:

- If they are all direction-consistent, the node also holds other rows (columns) whose indices are the column indices of nonzeros in the rows used for evaluating them.
- If they are all direction-opposite, the node also holds other rows (columns) whose indices are the row indices of nonzeros in the columns used for evaluating them.
- If they are both direction-consistent and direction-opposite, the node also holds other rows (columns) whose indices contain the column indices of nonzeros in the rows used for evaluating the direction-consistente edges and the row indices of nonzeros in the columns used for evaluating the direction-opposite edges.

Based on degree-driven evaluation order, the multi-stage data partitioner for **cyclic** queries remains unchanged.

For the query with **constants**, the multi-stage RDF partitioner divides the RDF matrix for each traversal root based on the light query evaluation results. Each root is adjacent to constants, hence the multi-stage RDF partitioner gives priority to the root that obtains the minimal number of bindings. In addition, the first-stage partitioning is changed to partitioning the specific rows (columns) whose indices in $A$ correspond to the bindings of the root:

- If the root only has outgoing query edges, the specific rows corresponding to the bindings of the root are partitioned by rows into $N_p \times N_t$ parts based on the number of rows.
- If the root only has incoming query edges, the corresponding columns are partitioned by columns into $N_p \times N_t$ parts based on the number of columns.
- If the root has both outgoing and incoming query edges, the corresponding rows and columns are partitioned into $N_p \times N_t$ parts.

*Example 6.5.* Figure 6 presents the degree-driven multi-stage partitioner for the example RDF data (Figure 1a) and SPARQL query (Figure 2), where we set $N_p = 2$ and $N_t = 2$. The number of edge levels of the only $Root_0$ ($v_2$) is $L_0 = 2$, and $L = 2$

**First-stage partitioning**: The $0^{th}$-level query edges consist of direction-consistent and direction-opposite edges. Thus the $2^{nd}$ row of $A$ (row 2) is eliminated from the LSpM$_{CSR}$-stored $A$, and the $3^{rd}$ column of $A$ (column 3) is eliminated from the LSpM$_{CSC}$-stored $A$. Then the two matrices respectively stored in LSpM$_{CSR}$ and LSpM$_{CSC}$ formats are partitioned into $N_p \times N_t = 4$ parts, where each part has one row and one column. Each compute node holds 2 parts, and each GPU thread in the node evaluates the $0^{th}$-level query edges on one part. The details are as follows:

The $0^{th}$ compute node is assigned the rows 0 and 1 and the columns 0 and 1. In addition, an array $Au[2]$ is built to record the index of each row (column) on the node, as shown in Figure 6. Wherein the $0^{th}$ GPU thread of the node evaluates $\overrightarrow{v_2, v_1}$ on row 0 and evaluates $\overrightarrow{v_0, v_2}$ and $\overrightarrow{v_3, v_2}$ on column 0. The $1^{st}$ GPU thread executes query evaluation on the row 1 and column 1.

The $1^{st}$ compute node is assigned the rows 4 and 5 and the columns 4 and 5, and the array $Au[2]$ is recorded on the node. Wherein the $0^{th}$ GPU thread executes query evaluation on the row 4 and column 4. The $1^{st}$ GPU thread executes evaluation on the row 5 and column 5.

**Next-stage partitioning**: The only $1^{st}$-level edge $\overrightarrow{v_0, v_1}$ is direction-consistent, so the next-stage partitioning selects other rows in the LSpM$_{CSR}$-stored $A$ for evaluating the $1^{st}$-level edge. In addition, the $0^{th}$-level query edge $\overrightarrow{v_0, v_2}$ has the common vertex $v_0$ with the $1^{st}$-level edge, and $\overrightarrow{v_0, v_2}$ is direction-opposite. Thus the indices of the other rows selected by the next-stage partitioning correspond to the row indices of the nonzeros in the columns used for evaluating $\overrightarrow{v_0, v_2}$. The details are as follows:

For the $0^{th}$ compute node, the row indices of the nonzeros in columns 0 and 1 used for evaluating $\overrightarrow{v_0, v_2}$ are 0, 1, 2, and 5. The node already holds the rows 0 and 1 using the first-stage partitioning, thus the node also holds the rows 2 and 5 in the next-stage partitioning. The arrays $Ir[8]$ and $Ic[8]$ are built to respectively index all the rows and columns hold by the node, where $Ir[i]$ ($Ic[i]$) is the index of row $i$ (column $i$) in all the rows (columns) held by the node.

For the $1^{st}$ compute node, the row indices of the nonzeros in columns 4 and 5 used for evaluating $\overrightarrow{v_0, v_2}$ are 2, 4, 6, and 7. While the node already holds row 4, and the rows 6 and 7 do not exist in the LSpM$_{CSR}$-stored matrix, thus the node further holds row 2. The arrays $Ir[8]$ and $Ic[8]$ are built to respectively index all the rows and columns hold by the node.

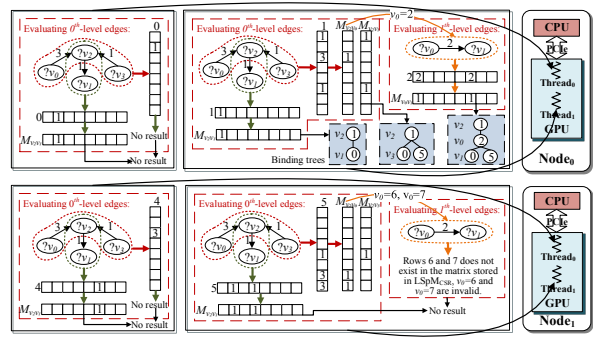

**Figure 7: The main computation phase of the example degree-driven gSmart (Figure 6).**

## 7 MAIN COMPUTATION PHASE

This section introduces two key techniques in the main computation phase of gSmart, including tree-based binding storage and fine-grained processing scheme.

### 7.1 Tree-based Binding Storage

According to Eqs. (13), (16), (20), and (24), there is a binding matrix with the same size as the corresponding RDF matrix after evaluating each query edge using the matrix algebra operators. To reduce memory consumption and simplify the final join operation, we propose a tree-based storage to maintain bindings of variables, i.e., nonzeros in binding matrices.

Given a query graph $G_q = (E, V)$, we form binding trees based on traversal paths generated by the traversal strategies (§ 6.1). Once each branch of the $Root_r$ has been traversed (before backtracking), the branch is recorded as a path. Binding trees are formed for each path. The root node at the first level (level 0) of a binding tree stores a binding of the $Root_r$. The nodes at each of the other levels (level $i$) store bindings of the $i^{th}$ vertex in the path, where these bindings are obtained based on the binding stored in the root node. Therefore, the number of binding trees for the path is the number of bindings of $Root_r$, and the number of levels of a binding tree is the number of vertices in the corresponding path.

*Example 7.1.* Consider the example SPARQL query (Figure 2) based on the degree-driven traversal, there are 3 paths of $Root_0$: "$v_2 \rightarrow v_1$", "$v_2 \rightarrow v_3$", and "$v_2 \rightarrow v_0 \rightarrow v_1$". As shown in Figure 7, the corresponding binding trees are only obtained by the $1^{st}$ GPU thread of the $0^{th}$ node.

### 7.2 Parallel Executor

Based on the partitioning obtained as described in § 6.3, the parallel executor first reads the matrix data obtained by the multi-stage partitioner into the host memory, and then transfers the data from host to device memory. Each GPU thread evaluates the given query a row- or/and a column-at-a-time. When the parallel executor completes, the binding trees are returned from device to host memory. Algorithms 1 and 2 describe gSmart parallel executor on each GPU thread.

**Algorithm 1** gSmart on each thread.

**Require:** The RDF matrix data obtained by the multi-stage partitioner;
**Ensure:** Binding trees.
1: **for** each $Root_r$ of the query graph **do**
2:    **for** each row or/and each column in the matrix data obtained in the first-stage partitioning **do**
3:       Evaluate query edges $\overrightarrow{Root_r, W_1}$ or/and $\overrightarrow{W_1, Root_r}$;
4:       **if** evaluation results exist **then**
5:          Form the binding sub-trees of $Root_r$ and $w_1$, where $w_1$ in $W_1$ is the end vertex in the corresponding path;
6:          $MAQP(W_1)$;
7: **return** Binding trees.

**Algorithm 2** The function $MAQP()$.

**Require:** $W_l$;
**Ensure:** Binding trees.
1: **for** each $w_l$ in $W_l$ that is not the end vertex in any path of $Root_r$ **do**
2:    $flag = 0$;
3:    **for** each binding of $w_l$ **do**
4:       Evaluate $\overrightarrow{w_l, W_{l+1}}$ or/and $\overrightarrow{W_{l+1}, w_l}$;
5:       **if** results exist **then**
6:          $flag = 1$;
7:          Form the binding sub-trees of $w_l$ and $w_{l+1}$, where $w_{l+1}$ in $W_{l+1}$ is the end vertex in the corresponding path;
8:          **return** $MAQP(l + 1)$;
9:    **if** $flag = 0$ **then**
10:      Delete the sub-trees that are formed based on the current binding of $w_{l-1}$.
11:      **break**;
12:   **if** $flag = 1$ **then**
13:      Insert the bindings of $w_{l-1}$ into the sub-trees if $w_l$ is not the end vertex in the corresponding path;
14: **return** Binding trees.

*7.2.1 Fine-grained processing scheme.* As demonstrated in Algorithms 1 and 2, the eligible incident query edges connecting a vertex are evaluated on the RDF matrix data obtained by the multi-stage partitioner one row- or/and one column-at-a-time, based on one of the bindings of the vertex. We call this *fine-grained processing*.

Each GPU thread starts from evaluating the $0^{th}$-level query edges of $Root_r$ on each row or/and column in the matrix data obtained in the first-stage partitioning (lines 2-3 in Algorithm 1). Next, there is a nested loop with multiple layers in the function $MAQP()$ (Algorithm 2). At each layer of the nested loop, select each vertex $w_l$ from the set $W_l$ that contains the common vertices in the $(l-1)^{st}$-level edges that have been just evaluated and the $l^{th}$-level edges that are eligible to be evaluated ($l \in \{1, 2, \ldots, L_r - 1\}$) (line 1 in Algorithm 2). In subsequent, select each binding of $w_l$, and evaluate the eligible $l^{th}$-level edges connecting $w_l$ on the row or/and column corresponding to the binding (lines 3-4 in Algorithm 2).

*7.2.2 Pre-pruning Technique.* To avoid unnecessary query processing overhead, two key timings are important for gSmart to prune invalid bindings:

(1) Prior to the start of evaluation for each query edge.

- If the evaluation for the $0^{th}$-level query edges of $Root_r$ generates no result, this indicates that no result exists in the current row or/and column; Thus the fine-grained processing scheme will stop evaluating the query edges at the next level, while try the next row or/and column in the matrix data immediately (lines 2 and 4 in Algorithm 1).
- If the evaluation for the $l^{th}$-level query edges of $Root_r$ connecting $w_l$ generates no result, this indicates that the current binding of $w_l$ is invalid, and the scheme will try the next binding of $w_l$ immediately (lines 3 and 5 in Algorithm 2).
- If the evaluation for the $l^{th}$-level query edges based on all the bindings of $w_l$ generates no result, this indicates that the current binding of $w_{l-1}$ is invalid; Thus the scheme will try the next binding of $w_{l-1}$ immediately. In partitcular when $l = 1$, $w_{l-1}$ is $Root_r$ and the scheme will try the next row in the matrix data (lines 9 and 11 in Algorithm 2).

(2) While binding trees are formed.

- If the evaluation for the $l^{th}$-level query edges connecting $w_l$ and $w_{l+1}$ generates results and $w_{l+1}$ is the end vertex in the path, the binding sub-trees storing bindings of $w_l$ and $w_{l+1}$ are formed (lines 5-7 in Algorithm 2).
- If the evaluation for the $l^{th}$-level query edges based on all the bindings of $w_l$ generates no result, all the binding sub-trees that are formed based on the current binding of $w_{l-1}$ are deleted (lines 9-10 in Algorithm 2).
- If the evaluation for all the quey edges based on the current binding of $w_{l-1}$ generates results, this binding of $w_{l-1}$ is inserted into the corresponding binding sub-trees (lines 12-13 in Algorithm 2).

*Example 7.2.* Figure 7 presents the main computation phase of degree-driven gSmart for the example RDF data (Figure 1a) and SPARQL query (Figure 2).

Each thread first evaluates the $0^{th}$-level query edges of $Root_0$ ($v_2$) on the only row and column obtained in the first-stage partitioning: evaluating $\overrightarrow{v_2, v_1}$ on the row, and evaluating $\overrightarrow{v_0, v_2}$ and $\overrightarrow{v_3, v_2}$ on the column (lines 2-3 in Algorithm 1). If the evaluation results exist, a binding tree of $v_2$ and $v_1$ and a binding tree of $v_2$ and $v_3$ are formed; Otherwise, the execution is terminated (lines 4-5 in Algorithm 1). Then based on all the bindings of $v_0$, the thread evaluates the $1^{st}$-level query edges: evaluating $\overrightarrow{v_0, v_1}$ on the corresponding rows held by the compute node (lines 1-4 in Algorithm 2). If the evaluation results exist, a binding tree of $v_2$, $v_0$, and $v_1$ is formed (lines 5-7 and 12-13 in Algorithm 2). Otherwise, the binding tree of $v_2$ and $v_1$ and the binding tree of $v_2$ and $v_3$ are deleted, and the execution is terminated (lines 9-11 in Algorithm 2). The details are as follows:

For the $0^{th}$ thread of the $0^{th}$ node, the evaluation for $\overrightarrow{v_0, v_2}$ and $\overrightarrow{v_3, v_2}$ in the $0^{th}$-level query edges generates no result, thus the execution is terminated immediately.

For the $1^{st}$ thread of the $0^{th}$ node, the evaluation for all the query edges yields results. Therefore, the three binding trees are obtained.

For the $0^{th}$ thread of the $1^{st}$ node, the evaluation for $\overrightarrow{v_0, v_2}$ and $\overrightarrow{v_3, v_2}$ in the $0^{th}$-level query edges generates no result, thus the execution is terminated.

For the $1^{st}$ thread of the $1^{st}$ node, the evaluation for the $1^{st}$-level query edge $(\overrightarrow{v_0, v_1})$ generates no result, thus the binding tree of $v_2$ and $v_1$ and the binding tree of $v_2$ and $v_3$ are deleted, and the execution is terminated.

## 8 POST-PROCESSING PHASE

The operations in the post-processing phase vary according to the query graph structure.
- For a conjunctive query graph with one traversal root:
  - If it has cycles or multiple constants, **local tree-pruning technique** (described below) is used to prune invalid bindings on CPU of each compute node.
  - Otherwise, the evaluation results obtained in the main computation phase satisfy all the constraints and are the final results without further computation.
- For a conjunctive query graph with more than one root:
  - If it has no constant:
    * If it is acyclic, **global tree-pruning technique** (described below) is used on CPU of the main compute node, known as the main MPI process.
    * If it is cyclic, **local tree-pruning** is first performed on each CPU, and then the **global tree-pruning** is performed on the CPU of the main compute node.
  - If it has constants:
    * If it has cycles or multiple constants, **local tree-pruning** is required.
    * Otherwise, no further computation is required.

### 8.1 Local Tree-pruning Technique

Local tree-pruning filters out from the binding trees that store the same binding of a traversal root invalid bindings of each common variable in different paths. This is performed on the CPU of each local node. Therefore, the resulting binding trees that store the same binding of the root also store the same bindings of the common variables. All the nodes perform local tree-pruning in parallel.

Let $\Omega$ be the set of common variables (except for the root) in different paths of a root. The local tree-pruning proceeds according to the following steps until there are no unprocessed variables in $\Omega$.

(1) Select an unprocessed variable $v$ from $\Omega$.
(2) Search for target nodes in the binding trees that store the bindings of $v$ and the same binding of the root, where target nodes store the bindings of $v$ and these bindings of $v$ do not exist in all the searched binding trees. If the target nodes exist, continue; Otherwise, go back to step 1.
(3) Remove all the sub-trees of target nodes.
(4) Remove the parent node of the last removed node if the parent node has only one child node until there is no parent node needs to be removed.

Given a **cyclic** query, $\Omega$ also contains the common variables that form cycles. Given a query with **multiple constants**, $\Omega$ also contains the variables that are adjacent to the constant vertices, and when the unprocessed variable $v$ selected from $\Omega$ is adjacent to the constants, the local tree-pruning finds target nodes in all the binding trees that store the bindings of $v$.

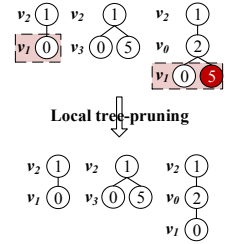

Figure 8: The post-processing phase of the example gSmart shown in Figure 7.

### 8.2 Global Tree-pruning Technique

The binding trees obtained by all compute nodes are gathered at the main node using MPI. Global tree-pruning, then, is executed on the CPU of the main node, and includes two parts: first invalid bindings of each common variable of different roots are filtered-out, then local tree-pruning (§ 8.1) is executed to update binding trees.

Let $\Phi$ be the set of common variables of different roots. Steps 1 and 2 of performing the first part of global tree-pruning are different from that of the local tree-pruning:

(1) Select an unprocessed variable $v$ from the $\Phi$ variables;
(2) Search for target nodes in all the binding trees that store bindings of $v$, where the bindings of $v$ are stored in the target nodes and these bindings of $v$ do not exist in all the searched binding trees of all the different roots; If the target nodes exist, continue; Otherwise, go back to step 1.

In addition, the common variables $\Omega$ in the second part of main-process tree-pruning are the common variables in different paths of the root.

*Example 8.1.* Figure 8 presents the post-processing phase of the example shown in Figure 7, where the main computation phase produces three binding trees storing the same binding of the $Root_0$ at the $0^{th}$ compute node. The query graph has only one traversal root and a cycle, which requires local tree-pruning on the local CPU to obtain the final results. The common variable that forms the cycle is $v_1$, thus steps 1 and 2 of the local tree-pruning technique search the first and third trees and find the target node storing the binding 5 of $v_1$, where the binding 5 of $v_1$ does not exist in the level that storing the bindings of $v_1$ of the first tree. Steps 3 and 4 delete the target node in the third tree.

## 9 EVALUATION

*Hardware setup.* gSmart is implemented in C. Single machine experiments are performed on one Intel(R) Xeon(R) CPU E5-2680 v4 @ 2.40GHz (14 cores each) with 256GB RAM and one NVIDIA Tesla P100 GPU with 16GB HBM2 stacked memory. Distributed experiments are conducted on Tianhe-1A supercomputer with up to 16 compute nodes. Tianhe-1A is a heterogeneous HPC system that combines CPUs and many-/multi-core accelerators. Each compute node has two Intel Xeon X5670 CPUs and one Nvidia Tesla M2050

Table 1: Details of the datasets, where #S&O is the number of unique subjects and objects, and #P is the number of unique predicates.

| Dataset | #Triples ($N$) | #S&O ($N$) | #P | Size (GB) |
|---|---|---|---|---|
| WatDiv-100M | 109.23 | 10.28 | 85 | 15 |
| YAGO2 | 284.30 | 60.70 | 98 | 42 |
| LUBM-1B | 1366.71 | 336.51 | 18 | 224 |

GPU; the CPUs are well suited to managing and serial processing, while the GPUs are good at parallel intensive processing. Each CPU and GPU have their private memory, respectively called host memory and device memory, connected via PCIe for data swapping, wherein the device memory has relatively small capacity while higher bandwidth compared to the host memory.

*Datasets.* We test gSmart on two synthetic and one real datasets widely used in a variety of work [25, 30, 48]. The details of the three datasets are listed in Table 1. WatDiv-100M with 109M triples is generated by the Stream WatDiv benchmark [5]. YAGO2 with 284M triples is extracted from Wikipedia[1], Geonames[2], and WordNet[3] [24]. LUBM-1B with 1367M triples is generated by the Lehigh University Benchmark (LUBM) [16]. The first two datasets presented in Table 1 are tested in single machine experiments, and the last two relatively large datasets are tested in the distributed experiments.

We use 20 benchmark SPARQL queries of four classes with varying characteristics: linear (L), star (S), snowflake (F), and complex queries (C) to test gSmart on WatDiv-100M dataset. Except for queries C1 and C3, all the other tested queries contain constants and only use the degree-driven traversal to pick the evaluation order of query edges. L, S, and F queries and query C2 all have only one traversal root. Query C1 has ?$v_0$ and ?$v_7$ as the roots of the direction-driven traversal, and ?$v_0$ is the only root of the degree-driven traversal. In query C3, the orders of query edges that are picked based on direction- and degree-driven traversals are the same: ?$v_0$ is the traversal root.

We use 9 queries to test on YAGO2, wherein the 4 benchmark queries (Y1-Y4) are defined in [1], and the other 5 queries (Y1$_c$, Y2′, Y2′$_c$, Y3$_c$, and Y4$_c$) are defined by modifying the 4 benchmark queries to extensively test gSmart on YAGO2. None of the 4 benchmark queries have constants, and queries Y1$_c$, Y2′$_c$, Y3$_c$, and Y4$_c$ have constants. Queries Y1, Y1$_c$, Y2, Y2′, Y2′$_c$, Y4, and Y4$_c$ are cyclic. The edge picking orders based on direction- and degree-driven traversals are the same and with a single traversal root for all the queries except for query Y3. In query Y3, ?$a1$ and ?$a2$ are the two roots of the degree-driven traversal, and either ?$a1$ or ?$a2$ can be the (single) root of the degree-driven traversal.

We use 7 benchmark queries (L1-L7) that are widely used in literature [1, 25, 38] to test on LUBM-1B. All the queries have constants and use the degree-driven traversal in gSmart with only one traversal root. Queries L1, L2, L3, and L7 are classified as data-intensive queries, and L4, L5, and L6 are classified as selective queries with relatively simple constructs and a small number of intermediate results.

*Comparative evaluation.* We compare gSmart against four state-of-the-art RDF query engines, including RDF-3X [30] (relation-based), gStore [48] and Wukong [38] (graph-based), and MAGiQ [25] (matrix-based). RDF-3X and gStore are single-threaded engines. Wukong is a multi-threaded and RDMA-enabled distributed-memory engine. MAGiQ supports single-thread (GraphBLAS), multi-thread (Matlab-CPU), single GPU (Matlab-GPU), and distributed-memory (CombBLAS) experiments. However, the complete source code of MAGiQ has not yet been released, so we can only present the single- and multi-thread experimental results of MAGiQ (GraphBLAS) and MAGiQ (Matlab-CPU).

### 9.1 Single Machine Performance

*9.1.1 Data Loading Overhead.* Table 2 presents the geometric mean of data loading overhead of gSmart for all the benchmark queries on a single machine. gSmart's data loading overhead on a single machine refers to the time spent on LSpM storage system, including the time of reading the raw datasets (step 1 of the LSpM storage system), encoding (step 2), storing the RDF data in LSpM format (steps 3 and 4 of the direction-driven LSpM system and steps 3-5 of the degree-driven LSpM system). gSmart has lower data loading overhead than other engines except MAGiQ, because gSmart eliminates unnecessary RDF triples and only reads and processes the necessary data in the compressed RDF matrix. We further analyze the loading overhead for each dataset below.

*Data loading overhead for WatDiv-100M.* Table 3 shows the breakdown of gSmart data loading overhead for WatDiv-100M based on different traversal methods, where "Read-Direction", "Encode-Direction", and "LSpM-Direction" respectively present the runtime of the direction-driven LSpM storage system only for queries C1 and C3, and "Read-Degree", "Encode-Degree", and "LSpM-Degree" respectively present the geometric mean runtimes of the degree-driven storage system for all the tested queries.

Table 2: Data loading overhead (min) comparison on a single machine.

| Dataset | RDF-3X | gStore | Wukong | MAGiQ | gSmart-Direction | gSmart-Degree |
|---|---|---|---|---|---|---|
| WatDiv-100M | 18 | 40 | 4 | **1** | 3 | 2 |
| YAGO2 | 78 | 63 | 9 | **3** | 3 | 3 |

Table 3: Breakdown of gSmart data loading overhead (s) for WatDiv-100M.

| Data Loading | L1-L5 | S1-S7 | F1-F5 | C1 | C2 | C3 |
|---|---|---|---|---|---|---|
| Read-Direction | - | - | - | 74.7 | - | 72.2 |
| Read-Degree | 70.0 | 71.5 | 73.9 | 74.7 | 75.3 | 72.2 |
| Encode-Direction | - | - | - | 20.9 | - | 258.4 |
| Encode-Degree | 6.7 | 12.5 | 18.0 | 20.9 | 39.3 | 258.4 |
| LSpM-Direction | - | - | - | 1.0 | - | 11.9 |
| LSpM-Degree | 0.6 | 1.0 | 1.8 | 2.0 | 4.6 | 11.9 |

---

[1]https://www.wikipedia.org/
[2]http://www.geonames.org/
[3]https://wordnet.princeton.edu/

**Table 4: Breakdown of gSmart data loading overhead (s) for YAGO2.**

| Data Loading | Y1 | Y2 | Y3 | Y4 | Y1$_c$ | Y2′ | Y2′$_c$ | Y3$_c$ | Y4$_c$ |
|---|---|---|---|---|---|---|---|---|---|
| Read-Direction | 195.7 | 199.1 | 187.6 | 187.2 | - | 199.1 | - | - | - |
| Read-Degree | 195.7 | 199.1 | 187.6 | 187.2 | 195.7 | 199.1 | 199.1 | 187.6 | 187.2 |
| Encode-Direction | 8.3 | 8.1 | 9.4 | 10.3 | - | 8.1 | - | - | - |
| Encode-Degree | 8.3 | 8.1 | 9.4 | 10.3 | 8.3 | 8.1 | 8.1 | 9.4 | 10.3 |
| LSpM-Direction | 0.4 | 0.4 | 0.4 | 0.4 | - | 0.4 | - | - | - |
| LSpM-Degree | 0.4 | 0.4 | 0.8 | 0.4 | 0.8 | 0.4 | 0.8 | 0.8 | 0.8 |

**Table 5: Geometric mean of execution time (ms) for WatDiv-100M.**

| Engine | L1-L5 | S1-S7 | F1-F5 | C1-C3 |
|---|---|---|---|---|
| RDF-3X | 11 | 11 | 32 | 813 |
| gStore | 230 | 139 | 187 | 1154 |
| Wukong | 1 | 16 | 2 | 47 |
| MAGiQ (GraphBLAS) | 790 | 1028 | 2168 | 5393 |
| MAGiQ (Matlab-CPU) | 16 | 25 | 44 | 234 |
| gSmart-Direction-CPU | - | - | - | 242 |
| gSmart-Direction-GPU | - | - | - | 96 |
| gSmart-Degree-CPU | **0.1** | **0.4** | **1** | 82 |
| gSmart-Degree-GPU | 1 | 1 | 6 | **41** |

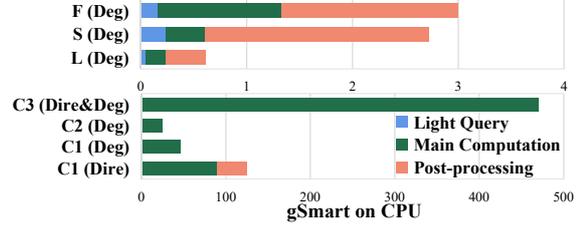
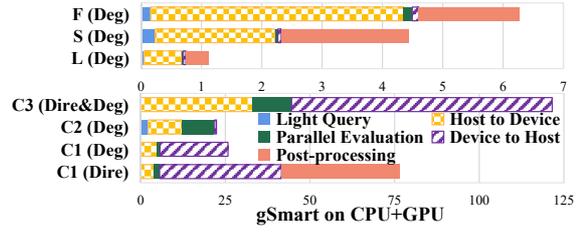

Figure 9: Breakdown of gSmart average runtime (ms) for WatDiv-100M.

The overhead of LSpM storing is much lower than reading the raw dataset and encoding. For all the query classes, the overhead of each part increases as the query structure becomes more complex. The costs of steps 1 and 2 of LSpM storage system remain the same regardless of the traversal strategy; therefore, the "Read-Direction" and "Encode-Direction" times are respectively the same as the "Read-Degree" and "Encode-Degree" times for queries C1 and C3. The "LSpM-Direction" time is around half the "LSpM-Degree" time for query C1, since direction-driven gSmart only stores the RDF data in LSpM$_{CSR}$ format, while direction-driven gSmart stores the RDF data in both LSpM$_{CSR}$ and LSpM$_{CSC}$ formats. The "LSpM-Direction" time is the same as "LSpM-Degree" for query C3, because all the query edges are direction-consistent, and only the LSpM$_{CSR}$ format is stored in both direction- and degree-driven gSmart. Degree-driven gSmart that shows the performance on all the queries loads this dataset slower than MAGiQ because the former may store the matrix twice (in both LSpM$_{CSR}$ and LSpM$_{CSC}$ formats) and spend double the amount of time on LSpM storing.

*Data loading overhead for YAGO2.* As shown in Table 4, the same edge picking orders result in the same data loading overhead based on different traversals for all the 4 benchmark queries except for query Y3. However, "LSpM-Deg" takes about twice as long as "LSpM-Dire" due to the extra overhead of LSpM$_{CSC}$ storing. Queries Y1-Y4 and Y2′ have no constant. Queries Y1$_c$, Y2′$_c$, Y3$_c$, and Y4$_c$ are created by modifying a variable respectively in queries Y1, Y2′, Y3, and Y4 into a constant; thus the direction-driven gSmart can only be used for queries Y1-Y4 and Y2′, and "Read-Deg" and "Encode-Deg" times for queries Y1$_c$, Y2′$_c$, Y3$_c$, and Y4$_c$ are respectively the same as Y1, Y2′, Y3, and Y4. However, "LSpM-Deg" times for Y1$_c$, Y2′$_c$, and Y4$_c$ are about double for Y1, Y2′, and Y4, respectively. This is because the light query evaluation for query Y1$_c$, Y2′$_c$, and Y4$_c$ processes direction-opposite edges that connect constants, and the LSpM storage system spends more times to store LSpM$_{CSC}$ format.

*9.1.2 Query Evaluation for WatDiv-100M.* As shown in Table 5, gSmart has significant performance superiority over all the competitors for WatDiv-100M. The light query evaluation prunes a large number of invalid bindings of the root for all queries except for C1 and C3, thus serial gSmart on CPU (gSmart-Degree-CPU) performs the best for L queries with the speedup of 10.00× (compared to Wukong) to 7900.00× (compared to MAGiQ (GraphBLAS)), for S queries with the speedup of 27.50× (compared to RDF-3X) to 2570.00× (compared to MAGiQ (GraphBLAS)), and for F queries with the speedup of 2.00× (compared to Wukong) to 2168.00× (compared to MAGiQ (GraphBLAS)). Due to the incident edge-based evaluation and efficient executor, gSmart on CPU+GPU (gSmart-Degree-GPU) achieves the optimal performance for C queries with the speedup up to 131.54×.

*Query workload evaluation.* Figure 9 presents the breakdown of average gSmart runtime for each query class on WatDiv-100M. Label "Light Evaluation" in Figure 9 represents the runtime of light query evaluation that evaluates the edges connecting constant vertices on CPU. Labels "Host to Device" and "Device to Host" in Figure 9b represent the communication time between host memory and device memory. The light query evaluation and post-processing are executed on CPU, thus "Light Query" and "Post-processing" runtimes in gSmart-CPU are the same as gSmart-GPU.

Table 6: Execution time (ms) for YAGO2.

| Engine | Y1 | Y2 | Y3 | Y4 | GeoMean |
|---|---|---|---|---|---|
| RDF-3X | 51 | 234600 | 9800 | 112 | 1904 |
| gStore | 274 | 136 | 8473 | 1053 | 758 |
| Wukong | **4** | 5 | 172 | 758 | 38 |
| MAGiQ (GraphBLAS) | 26069 | 33139 | 17331 | 21551 | 23834 |
| MAGiQ (Matlab-CPU) | 118 | 122 | 246 | 111 | 141 |
| gSmart-Direction-CPU | 34 | 34 | 123 | 26 | 44 |
| gSmart-Direction-GPU | 6 | **5** | 214 | **12** | 17 |
| gSmart-Degree-CPU | 34 | 34 | 79 | 26 | 39 |
| gSmart-Degree-GPU | 6 | **5** | **8** | **12** | **7** |

The "Light Query" and "Post-processing" runtimes for S and F queries are relatively long, because more than one edge in three of the seven S queries (S2, S4, and S5) and in two of the five F queries (F1 and F4) are evaluated in light query evaluation, and the local tree-pruning technique is required to filter out invalid bindings based on the results of main computation and the light query evaluation (queries S1, S3, S6, S7, F2, F3, and F5 have one constant, queries S2, S4, and S5 have two constants that are adjacent to the same variable vertex, and queries F1 and F4 have two constants that are adjacent to different variable vertices). No post-processing is required when degree-driven gSmart evaluates C queries, where all the C queries have one root and are acyclic, queries C1 and C3 have no constant, and query C2 has one constant.

As shown in Figures 9a and 9b, gSmart-CPU outperforms gSmart-GPU except for C queries, because the gSmart-CPU runtimes are very short, and the decrease in "Parallel Evaluation" time cannot offset the additional "Host to Device" and "Device to Host" runtimes in gSmart-GPU.

*Effect of query edge ordering.* The order in which edges are picked are different depending on the traversal methods only for query C1; we evaluate the effect of the two orders using this query. The rows of "C1 (Dire)" and "C1 (Deg)" in Figure 9a and Figure 9b show gSmart performance, respectively, based on direction- and degree-driven traversals.

As shown in Figure 9, degree-driven gSmart outperforms the direction-driven gSmart for query C1. Due to the efficient executor, the evaluation for each edge of query C1 in degree-driven gSmart is based on the bindings of an adjacent vertex. Thus unnecessary computation and invalid bindings are pruned, and the binding trees that are returned are the final results with the minimum size obtained under all constraints. However, query C1 in direction-driven gSmart has two roots, and the evaluation for edges of one root is not bound by the evaluation results of the query edges of the other root, which results in more computation and communication overhead. Therefore, compared to direction-driven gSmart for query C1, all parts except for "Host to Device" of degree-driven gSmart run faster, and the "Post-processing" part is additional in direction-driven gSmart to prune invalid bindings and generate the final results. Additionally, the "Host to Device" time of degree-driven gSmart is longer than direction-driven gSmart for C1, since degree-driven gSmart transfers more data stored in both LSpM$_{CSR}$ and LSpM$_{CSC}$ formats.

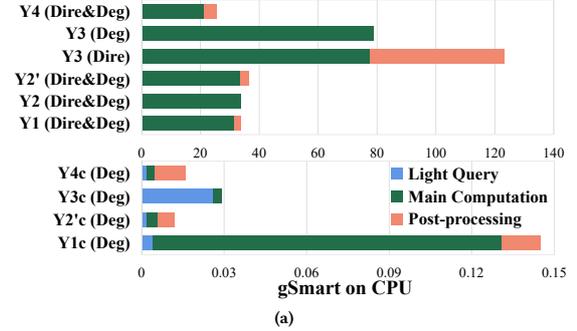
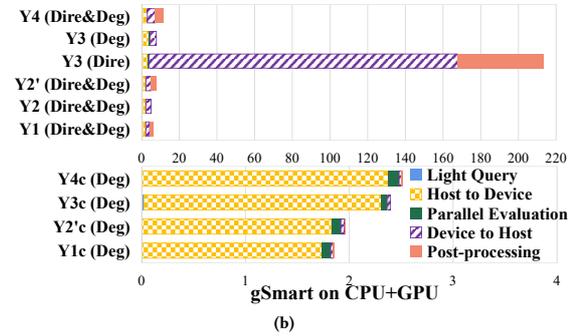

Figure 10: Breakdown of gSmart runtime (ms) for YAGO2.

*9.1.3 Query Evaluation for YAGO2.* The use of LSpM storage and efficiency of the execution engine minimizes the communication overhead; thus gSmart-GPU's performance is much better than other competitors for all the 4 benchmark queries except for query Y1, as shown in Table 6. The efficient parallelization in gSmart query evaluation contributes to this as well. While Wukong achieves the best performance for query Y1, gSmart-GPU is running a close second. gSmart-GPU and Wukong are tied for first place in the evaluation performance for query Y2 with the speedup of 24.40× (compared to MAGiQ (Matlab-CPU)) to 46920.00× (compared to RDF-3X). For query Y3, gSmart-Degree-GPU is the best-performing with the speedup of 21.50× (compared to Wukong) to 2166.38× (compared to MAGiQ (GraphBLAS)). gSmart on CPU+GPU works the best for query Y4 with the speedup of 9.25× (compared to MAGiQ (Matlab-CPU)) to 1795.92× (compared to MAGiQ (Graph-BLAS)). The geometric mean of runtimes of gSmart-Degree-GPU is the shortest with the speedup of 5.43× (compared to Wukong) to 3404.86× (compared to MAGiQ (GraphBLAS)), followed by gSmart-Direction-GPU.

*Query workload evaluation.* Figure 10 presents the breakdown of gSmart runtime for YAGO2, respectively, on CPU (Figure 10a) and CPU+GPU (Figure 10b). There is no "Light Query" runtime for queries Y1-Y4 and Y2′, so the computation and communication volumes without the selection by constant vertices are relatively large. The main computation phase of gSmart for query Y2 has no result (its "Post-proccessing" runtime is 0), hence query Y2′ is created by modifying two edges in query Y2 to further test the "Post-proccessing" runtime. In addition, no post-processing is required in degree-driven gSmart for queries Y3 and $Y3_c$ according to § 8.

**Table 7: gSmart data loading overhead (s) for YAGO2 on 16 nodes.**

| Data Loading | Y1 | Y2 | Y3 | Y4 | $Y1_c$ | $Y2'$ | $Y2'_c$ | $Y3_c$ | $Y4_c$ |
|---|---|---|---|---|---|---|---|---|---|
| Read | 220 | 239 | 244 | 235 | 220 | 239 | 239 | 244 | 235 |
| Encode | 32 | 33 | 33 | 39 | 32 | 33 | 33 | 338 | 39 |
| LSpM | 0.4 | 0.4 | 1 | 0.4 | 1 | 0.4 | 1 | 1 | 1 |
| Partition | 3E−2 | 2E−2 | 3E−2 | 2E−2 | 1E−2 | 2E−2 | 1E−2 | 3E−6 | 1E−2 |

**Table 8: gSmart data loading overhead (s) for LUBM-1B on 16 nodes.**

| Data Loading | L1 | L2 | L3 | L4 | L5 | L6 | L7 |
|---|---|---|---|---|---|---|---|
| Read | 1402 | 1547 | 1402 | 1851 | 1260 | 1400 | 2803 |
| Encode | 8767 | 11287 | 8765 | 9734 | 4264 | 5810 | 8641 |
| LSpM | 236 | 357 | 236 | 304 | 70 | 144 | 349 |
| Partition | 1 | 1 | 2 | 9E−6 | 0 | 1 | 3 |

Compared to other queries that have no constant, the "Main Computation", "Parallel Evaluation", and direction-driven "Post-proccessing" runtimes for query Y3 are the relatively longest. Because the numbers of $0^{th}$-level edges of each root of Y1 (4 edges of root ?p), Y2 and $Y2'$ (5 edges of ?p), and Y4 (3 edges of root ?p1) are more than that of Y3 (2 edges of root ?a1 and 2 edges of the root ?a2 in direction-driven gSmart, and 2 edges of root ?a1 (or ?a2) in degree-driven gSmart). The larger the number of the $0^{th}$-level edges, the more constraints are used to evaluate the rest of the query edges at other levels, and the less computation and intermediate results.

*Effect of query edge ordering.* Similar to WatDiv-100M, gSmart's performance based on the order in which query edges are picked can only be evaluated on query Y3. The effects of the ordering are shown in rows labeled "Y3 (Dire)" and "Y3 (Deg)" of Figure 10.

The "Host to Device" time of degree-driven gSmart is higher than direction-driven gSmart for query Y3. This is due to the fact that all the query edges are direction-consistent edges and the $LSpM_{CSR}$-stored matrix containing two predicates[4] is transferred in direction-driven gSmart, while degree-driven gSmart transfers the same $LSpM_{CSR}$-stored matrix as well as the $LSpM_{CSC}$-stored matrix containing one predicate[5] in the direction-opposite edges. Despite this, degree-driven gSmart significantly outperforms direction-driven gSmart for query Y3. This is because the binding trees returned in degree-driven gSmart are bounded by all the four query edges and are the final results with the minimal size, while in direction-driven gSmart, the binding trees of each root are bounded by only two query edges. Therefore, the direction-driven gSmart has a higher "Device to Host" time, as well as additional "Post-processing" time, which indicates a much larger number of binding trees are generated and an additional post-processing phase is required.

---
[4] http://yago-knowledge.org/resource/hasPreferredName and http://yago-knowledge.org/resource/actedIn
[5] http://yago-knowledge.org/resource/actedIn

### 9.2 Distributed Performance

According to evaluation results in single machine experiments, the degree-driven gSmart for YAGO2 achieves the better performance. In addition, all the 7 queries for LUBM-1B have constants and can only apply the degree-driven query plan. Thus, we only use the degree-driven query plan for all the distributed experiments, and there is no intra-node communication according to § 8.

*9.2.1 Data Loading Overhead.* Tables 7-8 present the breakdown of data loading overhead on 16 Tianhe-1A compute nodes respectively for YAGO2 and LUBM-1B. The geometric mean of data loading times for all the 9 queries on YAGO2 is approximately 4 minutes, for all the 7 queries on LUBM-1B it is approximately 161 minutes. The data loading overhead of gSmart in distributed environments adds the overhead of multi-stage partitioner, as shown as "Partition", where the multi-stage partitioner is used to exploit parallelism among compute nodes. However, the "Partition" runtime is relatively minor for other parts of data loading. The "Partition" runtimes for queries $Y3_c$, and L4 are very less because these two queries only have the first-level query edges. No "Partition" runtime for query L5 because the light query evaluation processes all the query edges (they are all incident to a constant vertex) and the multi-stage partitioner is not required.

*9.2.2 Query Evaluation.* Figure 11 shows the gSmart runtime for YAGO2 on multiple nodes. Scaling from 2 to 16 nodes, the average speedups of gSmart-CPU and gSmart-GPU are 6.66× and 2.90× for queries Y1-Y4 and $Y2'$ on YAGO2, 1.28× and 1.01× for queries $Y1_c$, $Y2'_c$, $Y3_c$, and $Y4_c$ on YAGO2.

Figure 12 shows the gSmart runtime for LUBM-1B on multiple nodes. Scaling from 2 to 16 nodes, the average speedups of gSmart-CPU and gSmart-GPU are 4.42× and 1.46× for data-intensive queries (queries L1, L2, L3, and L7) on LUBM-1B, and 2.63× and 1.44× for selective queries (queries L4, L5, and L6) on LUBM-1B.

gSmart has poor scalability for queries $Y1_c$, $Y2'_c$, $Y3_c$, and $Y4_c$ on YAGO2 because the light query evaluation generates a very small number of results for these queries, which limits the parallelism among compute nodes that the multi-stage partitioner of gSmart can exploit.

gSmart-CPU achieves the speedup up to 7.90× (for query Y1) on YAGO2 and 6.90× (for query L1) on LUBM-1B. gSmart-CPU is more scalable than gSmart-GPU on multiple nodes for the two datasets, and gSmart-GPU on YAGO2 performs better scalability than that on LUBM-1B. gSmart for data-intensive queries scales better than that for selective queries on LUBM-1B. The reason can be seen in Figure 10, "Host to Device" plays a large role in gSmart-GPU runtime. As the number of nodes increases, the size of arrays *Ir* or/and *Ic* remains unchanged (is N), while the transmission time for other arrays decreases. For large-dimensional RDF matrices having large N and selective queries involving less data, hence, the overhead of transferring *Ir* or/and *Ic* is an increasing percentage of the "Host to Device" time. This impairs the parallel scalability of gSmart-GPU on large-dimensional RDF data, especially for selective queries.

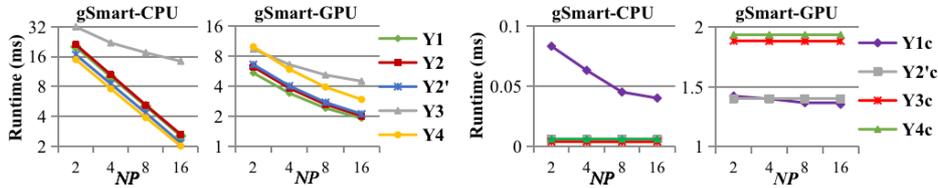

Figure 11: Runtime for YAGO2 scaling from 2 to 16 nodes.

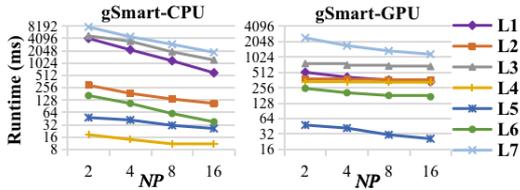

Figure 12: Runtime for LUBM-1B scaling from 2 to 16 nodes.

## 10 CONCLUSIONS

This paper introduces gSmart, an efficient SPARQL query engine using sparse matrix algebra, by leveraging the heterogeneous HPC architectures. By adopting grouped incident edge-based RDF query evaluation using sparse matrix algebra, gSmart can prune invalid results during query evaluation, reducing intermediate result sizes. A graph query planner is utilized by gSmart to determine the picking order of vertices in query graphs. Using the LSpM RDF storage and tree-based structure for bindings, gSmart minimizes the computation and communication costs. A multi-stage data partitioner is used by gSmart to exploit the hybrid and multi-level parallelism of the underlying architectures. Furthermore, a parallel executor is designed for gSmart to lower inter-node communication and enable high throughput. gSmart on a single CPU+GPU machine achieves the speedup of up to 46920.00× over the state-of-the-art, and achieves a maximum speedup of 6.90× scaling from 2 to 16 CPU+GPU nodes. In future work, we plan to further accelerate matrix algebra for the SPARQL query engine on heterogeneous HPC architectures.

## 11 ACKNOWLEDGMENTS

The research was partially funded by the National Key R&D Program of China (Grant No. 2018YFB0204302), the Programs of National Natural Science Foundation of China (Grant Nos. 61625202, 61806077), the International Postdoctoral Exchange Fellowship Program of China Postdoctoral Council (Grant No. OCPC2020063), and the Science and Technology Innovation Program of Hunan Province of China (Grant No. 2020RC2032). M. Tamer Özsu's research is funded by a Discovery Grant from the Natural Sciences and Engineering Research Council (NSERC) of Canada.